\documentclass[a4paper]{article}
\usepackage{ASVspoof2021}
\usepackage{multicol, blindtext}
\usepackage{epsfig,amssymb,amsmath}
\usepackage{stfloats} 
\usepackage{multirow}
\usepackage{hyperref}
\usepackage{pifont}
\usepackage{float}
\newcommand{\cmark}{\ding{51}}%
\newcommand{\xmark}{\ding{55}}%
\ninept

\title{A Study On Data Augmentation In Voice Anti-Spoofing}

\makeatletter
\def\name#1{\gdef\@name{#1\\}}
\makeatother

\name{{\em Ariel Cohen$^{*}$\thanks{ $^{*}$ equal contribution},  Inbal Rimon$^{*}$\footnotemark[1],  Eran Aflalo and Haim Permuter}}

\address{
Ben Gurion University, Be'er Sheva, Israel \\
{\small \tt ariel5, inbalri, eranaf@post.bgu.ac.il } \\
{\small \tt haimp@bgu.ac.il } 
}

\begin{document}
\maketitle

\begin{abstract}
In this paper we perform an in depth study of how data augmentation techniques improve synthetic or spoofed audio detection. Specifically, we propose methods to deal with channel variability, different audio compressions, different bandwidths and unseen spoofing attacks, which have all been shown to significantly degrade the performance of audio based systems and Anti-Spoofing systems. Our results are based on the ASVspoof 2021 challenge, in the Logical Access (LA) and Deep Fake (DF) categories. Our study is \emph{Data-Centric}, meaning that the models are fixed and we significantly improve the results by making changes in the data. We introduce two forms of data augmentation - compression augmentation for the DF part, compression \& channel augmentation for the LA part. In addition, a new type of online data augmentation, SpecAverage, is introduced in which the audio features are masked with their average value in order to improve generalization. Furthermore, we introduce a Log spectrogram feature design that improved the results. Our best single system and fusion scheme both achieve state of the art performance in the DF category, with an EER of 15.46\% and 14.46\% respectively. Our best system for the LA task reduced the best baseline EER by 50\% and the min t-DCF by 16\%. Our techniques to deal with spoofed data from a wide variety of distributions can be replicated and can help anti-spoofing and speech based systems enhance their results. \\ 
\textbf{Keywords}: ASVspoof 2021, Audio Data Augmentation, Data-centric AI, SpecAugment, Voice anti-spoofing, Voice Deep fake.

\end{abstract}

\section{Introduction}
The use of the human voice for tasks such as Automatic Speaker Verification (ASV), spreading news on social media, and communicating using digital devices has become very popular. ASV, for example, is used in many applications such as voice mail, telephone banking, call centers, biometric authentication, forensic applications and more. \par 
Generating synthetic speech has become a doable task now a days, as more and more algorithms emerge and technology advances. These algorithms include Text to Speech (TTS) \cite{TTS} , Voice Conversion (VC, converting speech from source speaker to target speaker), \cite{zhao2020voice,kobayashi2021crank}, among others. Spoofing, is the process of creating synthetic speech where the goal is either 
to fool algorithm-based solutions/automatic solutions or the human ear, by creating perceptually natural sounding speech that mimics a target speaker. Another form of spoofing can be physically replaying a recorded audio sample of a specific speaker. Research has shown that both Technology and the human ear are susceptible to voice spoofing. In the past few years, Anti-Spoofing for ASV has become a field of interest in the research community, as four challenges \cite{asvspoof2015,asvspoof2017,asvspoof2019,2021asvspoof} have been held in which the goal has been to improve the ability to discriminate bona fide speech from spoofed speech. \par

Aside from the challenges of detecting whether a given audio signal is bona-fide or spoofed, practical Anti-Spoofing systems face the following challenges:
\begin{enumerate}
    \item \textbf{Compression:} Audio compression (lossy) typically contains some form of non linear quantization together with selective frequency reduction. Compression can be a cause of audio quality degradation and transmission mismatch that can degrade the performance of audio systems such as ASV systems \cite{lossycompressionon_ASV}, speaker recognition systems \cite{lossycompression_speakerrecognition}, and Anti-Spoofing systems. Common compressions are MP3 \cite{mp3}, Advanced Audio Coding (AAC) \cite{aac} and G.722  \cite{G722}, among others. 
    
    \item \textbf{Channel effects:} Transmitting compressed audio through a channel might induce transmission related data loss such as packet loss, noise and more. This type of data loss can degrade the performance of audio feature based systems, as stated in \cite{packet_loss}. Channels for example can be VoIP, Landline, Cellular and Satellite.  
    
    \item \textbf{Bandwidth differences and filtering:} Audio codecs can differ by bandwidth as well, as some codecs are narrow band codecs and some codecs are wide band codecs. In addition, some include band pass filtering prior to transmission, a fact that can cause information loss of high frequencies, which may contain crucial information necessary to detect spoofing attacks \cite{tak2020explainability}.
    
    \item \textbf{Unseen spoof attacks:} One of the main challenges of a spoofing system is to be able to generalize and to detect unseen attacks from an unknown distribution. In \cite{zhang2021empirical}, the authors performed a cross dataset study that included the VCC2020  \cite{VCC2020} dataset among others and showed significant degradation in performance.

\end{enumerate}

\noindent As stated in the evaluation plan \cite{evaluationplan2021}, the ASVspoof 2021 challenge contained two scenarios that included the issues stated above. Both scenarios contained bonafide and spoofed speech segments that have been generated with TTS and VC algorithms. In the Deep Fake (DF) category, new and never seen before spoofing methods have been used, and the audio files might have undergone compression (such as MP3, m4a and others) with various bit rates. In the Logical Access (LA) category, the audio files have been communicated across telephony and VoIP networks with various coding and transmission effects. Both scenarios contained unseen spoofing attacks. We used these two scenarios to benchmark our ideas.


\subsection{Model-Centric vs. Data-Centric}
An Artificial Intelligence (AI) system is typically composed of data and a model, while both go hand in hand in producing the desired results. A normal optimization process consists of constantly improving the statistical model and the data in an iterative manner. While both are important, the attention usually shifts towards one of the following:
\begin{enumerate}
    \item \textbf{Model-Centric Approach:} In this approach, the data is fixed and empirical tests are performed with respect to the model architecture and training procedure in order to maximize the results.
    \item \textbf{Data-Centric Approach:} In this approach, the model is fixed and changes/improvements are constantly made in the data set in order to maximize the results.
\end{enumerate}
Our study is mainly Data-Centric. We chose models which had good performance on the ASVspoof 2019 data, and focused our efforts on data augmentation and feature design in order to tackle the challenges of ASVspoof 2021.  


\subsection{Motivation}
\begin{figure}[h!]
\includegraphics[width=8cm]{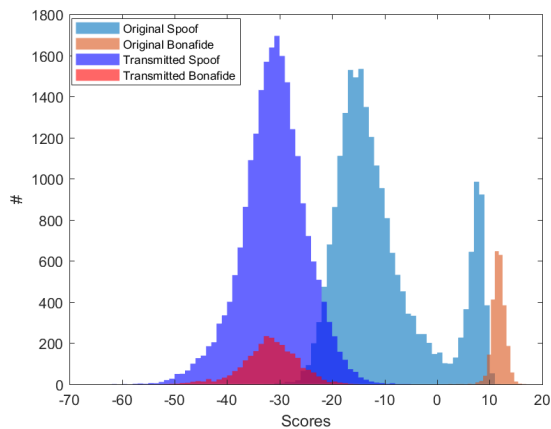} 
\caption{\it Score histograms from Resnet trained on the original training data set and tested on two development sets (original and transmitted). Each data set is separated to spoof and bonafide. The original data scores (blue and orange) are quite separable, while the augmented data is completely overlapping and indistinguishable (purple and red).}
\label{logspec_phenomena}
\end{figure}

In Figure \ref{logspec_phenomena}, we can see how the channel mismatches caused by compression, transmission effects, and bandwidth differences affect the score distribution of the Resnet model (\ref{ResNet}) with Log Spectrogram features. In this experiment, we trained the Resnet model using the original ASVspoof 2019 training set. Scores were produced on both the original ASVspoof 2019 development set, and a reference development set that has witnessed simulated transmission, possible packet loss and compression. Aside from the differences in the score range, we can see that the original data is relatively separable (bonafide scores are mostly different than spoofed scores), whereas the transmitted data is not separable. This led to a high EER both on our simulated transmitted data set and the evaluation data set. This demonstrates the sensitivity of audio based systems to harsh changes in the channel, and the importance of channel related data augmentation. We encountered similar effects using compressions. These findings motivated our work.

\subsection{Database and Setup}
In order to comply with the ASVspoof 2021 evaluation plan \cite{evaluationplan2021}, we used strictly the ASVspoof 2019 LA data set for training and development. The evaluation data was provided without labels by the ASVspoof 2021 organizers and scores were obtained using the challenge website during the post evaluation phase. Training and development partitions were kept the same as in ASVspoof 2019 for both the DF and LA parts of the competition. In this paper, data that hasn't been augmented is referred to as original data.

\subsection{Main Contributions}
In this paper we introduce several data augmentation techniques that improve the robustness of anti spoofing systems to channel variability and compression: 
\begin{enumerate}
    \item We introduce compression data augmentation methods that improve anti-spoofing systems performance with compressed data.
    \item We introduce channel robust data augmentation methods that improve anti-spoofing systems performance with compressed data that has been transmitted, filtered and down sampled.
    \item We introduce a new form of online data augmentation, SpecAverage, that is a variant of SpecAugment \cite{park2019specaugment}. In our experiments, SpecAverage has shown better performance in the tested scenarios. 
    \item We introduce two new feature design ideas: a log spectrogram feature design called double sided spectrogram centering, and online feature normalization. We show how each of these methods improve the robustness and performance in anti-spoofing systems.
    
    
    
\end{enumerate}

\noindent Our ideas were tested in the ASVspoof 2021 challenge in the Deep Fake (DF) category and in the Logical Access (LA) category: The use of our augmentation methods and new ideas resulted in state of the art performance in the deep fake category, both for our single system and for our system fusion.
\par The rest of the paper is arranged as following:
Section 2 contains the the deep learning models we used. In Section 3 we elaborate about the features we used and provide analysis regarding the effect that compression and transmission have on them. In section 4 we introduce new feature design ideas. In Section 5 we present our data augmentation methods: compression augmentation (for DF) and channel augmentation (for LA). We then introduce SpecAverage. Section 6 and 7 contain our results and analysis for the DF and LA parts. In section 8 we discuss insights obtained from our work, and section 9 contains conclusions and future work.
Our augmentation methods are publicly available at: https://github.com/InbalRim/A-Study-On-Data-Augmentation-In-Voice-Anti-Spoofing.



\section{Models}
In this section we present the Deep Learning models we used: Resnet-34, SEnet and One Class Softmax (OCS) Resnet.

\subsection{ResNet} \label{ResNet}
ResNet-34 is commonly used for image and audio tasks, as shown in \cite{senet}, \cite{resnet}. The architecture we used is based on Resnet as shown in \cite{senet}, with 2 main modifications based on empirical experiments conducted prior to this work:
\begin{enumerate}
    \item \textbf{Optimizer:} We used the AdamW \cite{adamw} optimizer. The AdamW optimizer includes different parameters. Weight decay (WD),  $\beta_{1}$ and $\beta_{2}$. We chose \(\text{WD} = 5\cdot10^{-8}\) and \(\beta_{1},\beta_{2} = (0.81,0.8991)\)
    \item \textbf{Loss Function:} We experimented with Binary Cross-Entropy (BCE) and Binary Focal-Loss (BFL) and with different class weights. We got the best results using class weights where the bona fide class is weighted 10 times more than the spoof class with both loss functions. In addition, BCE provied slightly better results than BFL.
\end{enumerate}

%
%

\subsection{SENet}
While in ResNet blocks the inputs channels are equal weighted, in SE blocks a different weight is given to each channel using Squeeze and Excitation (SE). As suggested in \cite{senet}, the squeeze is performed using average pooling. To simplify calculations, in order to perform excitation the input dimension is first reduced followed by a ReLU activation, and then extended followed by a sigmoid activation.

\subsection{OCS-ResNet}
In \cite{zhang2021one}, the authors 
presented a model based on Resnet-18 with an attentive pooling layer, and a one class softmax function as following:


\begin{equation}
    L_{OCS} = \frac{1}{N}\sum^{N}_{i=1}\log{(1+e^{\alpha(m_{y_{i}}-\hat{w}_{0}\hat{x}_{i})(-1)^{y_{i}}})}
\end{equation}
where $ m_{0},m_{1}\in[-1,1], m_{0}>m_{1} $ denote the angular margins between the classes, $\hat{w}_{0}$ denotes a normalized weight vector, $y_{i}\in \{0,1\}$ denotes the class and $\hat{x}$ denotes the normalized vector of a target class embedding. We used this model, with the same parameters as stated in the paper.

\subsection{Model Training}
Each one of the three models was trained separately, In order to achieve minimal correlation and maximize the fusion results. different augmentation methods were used for each model, both for training and development. All of the training and development data has been chosen without any additional knowledge about the evaluation set, aside from what is stated in the evaluation plan. The specific composure of the augmented data sets is stated in the upcoming sections. 





\section{Audio Processing and Features}
In this section we state the audio features we used with each model. We visualize and provide insights of how compressions, different channels and other conditions affect those audio features. 


\subsection{Log Spectrogram (LogSpec)}  \label{logspec}
The spectrogram of an audio signal has been proven to be effective as a neural network input in \cite{impact_of_audio_on_nn} and specifically for spoof detection in \cite{assert}. The spectrogram created using STFT (short-time fourier transform), is calculated as following:

\begin{equation}
\mathrm{LogSpec} = \log(|STFT(x)|^2)
\label{eq1}
\end{equation}

\noindent where \(x\) is the audio signal. We used a frame length of $25ms$, a hop of $10ms$, and a total fixed length of 5 seconds. LogSpec was used with the SEnet and Resnet models.

\subsection{Linear Frequency Cepstral Coefficients (LFCC)}
We used LFCC with a window size of $20ms$ and an overlap of $10ms$. We used both $\Delta$ and $\Delta\Delta$ as dynamic features. 20 coefficients were selected and pre emphasis was performed. The above resulted in a frequency dimension of 60, while the time dimension was fixed to 450. We used repeat padding for shorter utterances and randomly sliced the features out of longer utterances. LFCC was used with the OCS-ResNet model.

\subsection{Compression Effects on LFCC Features}
\begin{figure}[h!]
\includegraphics[trim={150pt 70pt 170pt 0},clip,width=\columnwidth]{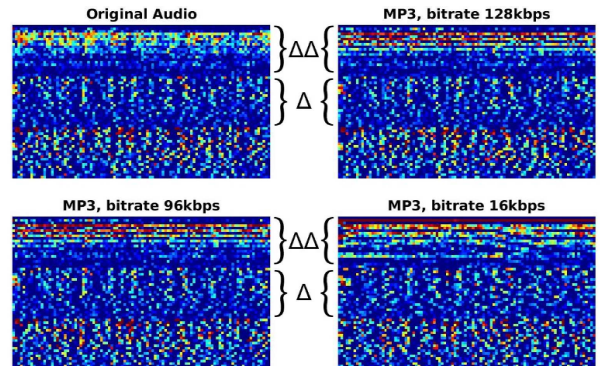} 
\caption{{\it LFCC's of compressed audio comparison. Compression affects the dynamic features ($\Delta$ and $\Delta\Delta$) significantly. }}
\label{LFCC_compression}
\end{figure}

In Figure \ref{LFCC_compression} we can see four images of LFCC's of the same utterance. Processing included MP3 compression with various bitrates and then decompression back to 16 kHz and 16 bits per sample FLAC format, to match the data format stated in the competition evaluation plan. As seen in Figure \ref{LFCC_compression}, MP3 compression affects the LFCC of a given input. While the static part (lower 20 coefficients) are affected mildly, the dynamic features ($\Delta$ and $\Delta\Delta$) are changed significantly, even between different bitrates of the same compression. 

\subsection{Compression and Transmission Effects on Audio Signals}
\begin{figure*}[h!]
\includegraphics[trim={110pt 0pt 90pt 0pt},clip, width=485pt]{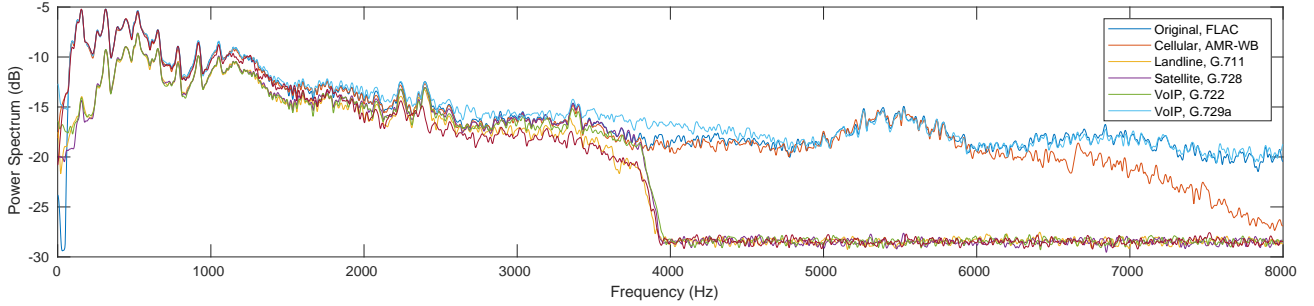} %
\caption{{\it Power spectrum of the same utterance, augmented with different transmissions and codecs. Differences are significant in the lower band (0 kHz - 1 kHz) where speech frequencies are located, mid band (3.5 kHz - 4.5 kHz) where the effect of downsampling can be seen, and upper band (7 kHz - 8 kHz). All emphasize the channel variation.}}
\label{power_spectrum}
\end{figure*}

\noindent In Figure \ref{power_spectrum}, we can see that for the same utterance, there are significant differences in the frequency response between the original audio file, and processed versions of the original file that have undergone compression, simulated transmission, packet loss, filtering and down sampling. All of the above are clearly a source of channel mismatches that degrade performance, as seen in Table \ref{LA_effect_compression}. 

\section{Feature Design}
We now introduce two feature design ideas that improved our results. We show how using a double sided spectrogram improves the results, and we then elaborate on our online normalization methods. 
\subsection{Double Sided Spectrogram}
Due to the fact that audio is a real signal, the Log spectrogram is symmetrical in frequency. And so, it seems that using the double sided LogSpec as a feature may not be useful, as it contains redundant data and might just cost run time. Despite that fact, motivated by the fact that we are using deep learning architectures that were heavily optimized to perform on images, we decided to re-test the use of double sided spectrograms. Images usually contain the object of interest in the center of the frame, and that motivated us to further explore that idea, since using a double sided Logspec would put the speech frequencies in the center of the frame. We choose to focus on two parameters: double side / one side and the resolution, which is set by the number of frequency bins / DFT points, nfft. nfft is typically chosen with respect to the frame length and sampling rate:

\begin{equation}
\text{nfft} = 2^{{\left\lceil\log(frame\_length \cdot fs)\right\rceil}}
\label{eq1}
\vspace{0.5cm}
\end{equation}

\noindent Where $fs$ is the sampling frequency (16 kHz in our case) and the frame length is $25ms$. The evaluation set is the DF evaluation set, that contains unseen spoof attacks, unseen compression methods and a variety of unknown bitrates.  We conducted a few experiments and achieved the following results:\\

\begin{table}[h!]
    \centering
    \begin{center}

\centerline{}
\begin{tabular}{|c|c|c|c|} 

 \hline
 Double Sided & nfft &  DF Eval EER \\
 \hline
 \xmark & 512 &   25.24 \\
 \hline
 \cmark & 512 &  \textbf{18.81} \\
 \hline
 \cmark & 1024  & 20.45 \\
 \hline

\end{tabular}
\caption{Double Sided LogSpec comparison, ResNet. A significant improvement can be seen using our feature design.}
\label{double_sided}
\end{center}
\end{table}

\begin{figure}[h!]
\includegraphics[width=\columnwidth]{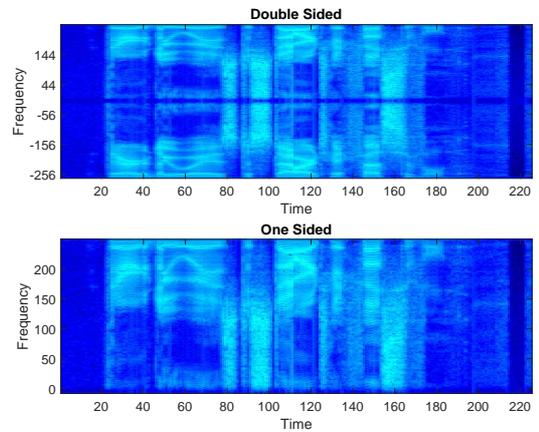} 
\caption{{\it Double sided and one sided LogSpec feature design. while no new information is added, the performance improves significantly. }}
\label{spectogram}
\end{figure}

\noindent As seen in Table \ref{double_sided}, using Resnet with the double sided LogSpec reduced the Deep Fake evaluation EER by 25.45\%. Increasing the resolution on the other hand, did not improve the performance as much and was incredibly costly in run time. This is an almost surprising result, as a mere centering of the input data improved the results significantly, without adding any additional information.  




\subsection{Feature Normalization}
One of the methods that had impressive empirical performances in the Deep Fake category was feature (LFCC / LogSpec) normalization. We experimented with 3 different types of normalizations:

\begin{centering}
\begin{equation}
\mathcal{N}_{\mathrm{min-max}}(x) = \frac{x-\min(x)}{\max(x)-\min(x)}
\label{01normalizations}
\end{equation}

\begin{equation}
\mathcal{N}_{\mathrm{mean}}(x) = \frac{x-\mathrm{mean(x)}}{\max(x)-\min(x)}
\label{mean_normalizations}
\end{equation}

\begin{equation}
\mathcal{N}_{\mathrm{standard}}(x) =  \frac{x-\mathrm{mean(x)}}{\mathrm{std}(x)}
\label{std_normalizations}
\end{equation}
\end{centering}

\noindent where $x$ denotes the feature matrix. All normalization methods are performed per individual sample, online (during training and testing). The computational price is cheap, as neither of the methods above affected our run time. We achieved the best performance in all systems using equation \ref{01normalizations}, where a simple linear projection mapped the features to $[0,1]$. The $\mathcal{N}_{\mathrm{min-max}}$ normalization reduced the OCS-ResNet Deep Fake EER by 18\%, and ResNets Deep Fake EER by 7\%, as stated in Table \ref{DF_singles}.




\section{Data Augmentation Techniques}

In this section we will introduce two forms of offline augmentation techniques that have been used to increase the performance of the anti spoofing systems with compressed data and channel variation. In addition, we will introduce SpecAverage, a new form of online data augmentation.



\subsection{Compression Augmentation} \label{compress augment}
The augmentation was performed as following:
\begin{enumerate}
    \item an audio file was chosen.
    \item a compression method was chosen out of: MP3 and AAC/m4a. 
    \item a bitrate was chosen.
    \item the audio file was compressed with the chosen method and bitrate.
    \item the compressed audio file was de-compressed back to FLAC format, 16 kHz, 16 bits per sample.
\end{enumerate}
\noindent The steps are performed to comply with the evaluation plan of ASVspoof 2021, in the DF part. We used a package called Pydub \cite{robert2018pydub} and FFmpeg \cite{tomar2006converting} in order to first read the audio files and then augment them. Table \ref{DF_data} contains the augmented part of our DF dataset. MP3 and AAC/M4a codecs were chosen since they were stated in the evaluation plan. The bitrates were chosen based on standard usages for each compression method, including high quality and low quality. This augmentation method was performed offline, since using it online and then extracting LFCC/LogSpec features was incredibly costly in CPU usage due to the feature extraction (and not the augmentation).



\begin{table}[h!]
    \centering
    \begin{center}

\begin{tabular}{|c|c|c|c|c|} 

 \hline
 Bitrate [kbps] & MP3 & AAC/m4a  \\
 \hline
 16 & \cmark &      \\
 \hline
 48 & \cmark &   \\
 \hline
 64 &   &\cmark  \\
 \hline
 96 & \cmark &\cmark  \\
 \hline
 128 &\cmark &\cmark   \\
 \hline
 160 &\cmark &    \\
 \hline
\end{tabular}
\caption{Deep Fake Compression Data Set}
\label{DF_data}
\end{center}
\end{table}






\noindent In Table \ref{EER_performance_DF} we see the effect that compression augmentation had on our models. A significant decrease in the EER can be seen after applying data augmentation.

\begin{table}[h!]
\vspace{2mm}
\centerline{
\begin{tabular}{|c|c|c|c|}
\hline
System & Augmentation &  DF Eval EER \\
\hline  \hline
OCS-resnet & \xmark &  29.31 \\
OCS-resnet & \cmark & \textbf{28.52} \\
\hline
Resnet & \xmark & 46.49 \\
Resnet & \cmark & \textbf{17.51} \\
\hline
SEnet & \xmark & 40.32 \\
SEnet & \cmark & \textbf{19.47} \\

\hline
\end{tabular}}
\caption{\it EER performance on compressed data (DF data set) with and without augmentation in the training and development sets.}
\label{EER_performance_DF}
\end{table}

\subsection{Codec, Channel Effect, Bandwidth Difference Augmentation}\label{codec augment}
In order to perform this augmentation we used the audio degradation simulator in \cite{ferras2016large} to augment the training and development data as following:
\begin{enumerate}
    \item an audio file was selected.
    \item for each Channel (Landline, Cellular, VoIP or Satellite) a random codec out of the list in Table \ref{LA_data_table} was chosen.
    \item RMS normalization was performed in order to simulate transmission gain changes. Normalization values were chosen from a uniform distribution from [-30,-10] in dB.
    \item Downsampling and band pass filtering were performed depending on the chosen codec.
    \item The audio file was compressed according to the chosen codec, with a random bitrate.
    \item Random packet loss was simulated.
    \item The audio file was re sampled to either 8 kHz  or 16 kHz depending on the data set we wanted to create.
\end{enumerate}

\begin{table}[h!]
    \centering
    \begin{center}
    
\begin{tabular}{|p{1cm}||p{1cm}|p{1.3cm}|p{1cm}|p{1cm}|} 
 
 \hline
 Channel & Landline & Cellular & VoIP & Satellite\\
 \hline
 \hline
 Codecs & G.711 G.726 & AMR  AMRWB  GSM  & Silk G.722 SilkWB G.729 & G.728  \\
 \hline

\end{tabular}
\caption{Logical Access data set channels and codecs used.}
\label{LA_data_table}
\end{center}
\end{table}

\noindent This type of augmentation was tested on the LA part. The channels and codecs we used are presented in Table \ref{LA_data_table}. Augmenting the data in this way, we kept the original train and development partitions, and created a data set for down sampled data (8 kHz), regular data (16 kHz) and wideband codecs data (16 kHz) as stated in Table \ref{LA_dataset_table}. In Table \ref{LA_effect_compression}, we can see that our augmentation improves EER and min t-DCF \cite{kinnunen2020tandem} significantly. All three models are tested on the LA evaluation data, with and without augmentation. 

\begin{table}[th]

\vspace{2mm}
\centerline{
\begin{tabular}{|c|c|c|c|}
\hline
System & Augmentation & Eval min t-DCF & Eval EER \\
\hline  \hline
OCS-resnet & \xmark & 0.7500 & 21.41 \\
OCS-resnet & \cmark & \textbf{0.3639} & \textbf{6.59} \\
\hline
Resnet & \xmark & 0.9032 & 40.73 \\
Resnet & \cmark & \textbf{0.2931} &  \textbf{5.18} \\
\hline
SEnet & \xmark & 0.9696 & 38.20 \\
SEnet & \cmark & \textbf{0.2961} &  \textbf{6.14} \\

\hline
\end{tabular}}
\caption{\it EER, min t-DCF performance on compressed data (LA data set) with and without augmentation.}
\label{LA_effect_compression}
\end{table}


\subsection{SpecAverage}


Replacing random blocks of the input feature with a constant value during the training process (Masking), was shown to force Deep Neural networks to be more robust and improved performance both in speaker recognition and in Anti-Spoofing, as shown in \cite{park2019specaugment} , \cite{chen2020generalization}. Inspired by that, We will first define general parameters and then elaborate about augmentation policies, that have been used both on LFCC and LogSpec features:
\begin{enumerate}
  \item \textbf{Frequency masking:} $f$ consecutive  frequency bins or coefficients $[f_{0}, f_{0} + f)$ are masked with a constant value, where f is chosen from a uniform distribution from 0 to the frequency mask parameter F, and $f_{0}$ is chosen from $[0, \nu - f)$. $\nu$ is the number of frequency bins or coefficients.
  \item \textbf{Time masking:} is applied so that $t$ consecutive time steps $[t_{0}, t_{0} + t)$ are masked with a constant value, where t is first chosen from a uniform distribution from 0 to the time mask parameter T, and $t_{0}$ is chosen from $[0, \tau - t)$. $\tau$ is the total number of time steps.
  \item \textbf{SpecAugment:} Masking with the value 0.
  \item \textbf{SpecAverage:} Masking with the average feature value calculated online for each feature.
\end{enumerate}

\noindent Keeping the same notation as in \cite{park2019specaugment}, Table \ref{online_augmentation_table} contains the training policies used:

\begin{table}[H]
\begin{center}
\centering
\begin{tabular}{|c|c|c|c|c|c|c|c|} 
 
 \hline
 Policy & Method & Feature & $m_{f}$ & $m_{t}$ & T & F\\
 \hline
 None & -  & - &  0 & 0 & 0 & 0   \\
 \hline
 SAv1 & SpecAverage & LFCC & 1 & 0 & 0 & 12   \\
 \hline
 SAu1 & SpecAugment & LFCC &  1 & 0 & 0 & 12   \\
 \hline
  SAv2 & SpecAverage & LFCC &  1 & 1 & 80 & 12 \\
 \hline
  SAv3 & SpecAverage & LogSpec &  0 & 1 & 10 & 0   \\
 \hline
 SAu3 & SpecAugment & LogSpec &  0 & 1 & 10 & 0   \\
 \hline
  SAv4 & SpecAverage & LogSpec &  1 & 0 & 0 & 10   \\
 \hline
 SAu4 & SpecAugment & LogSpec &  1 & 0 & 0 & 10   \\
 \hline

\end{tabular}
\caption{Augmentation policy. $m_{f}$ and $m_{t}$ denote the number of frequency and time masks respectively. T and F represent the amount of time values or frequency bins masked.}
\label{online_augmentation_table}
\end{center}
\end{table}

\noindent In Table \ref{DF_singles_SA} we can see that SpecAverage has boosted the performance of both ResNet and OCS-resnet with 2 different features. This leads us to believe that the method helps generalize and deal with unseen spoofing attacks and unseen compressions. It is important to note that SpecAugment has been implemented without time warping, which was computationally costly. The comparison we did was between masking with the value 0 and masking with the average value.

\begin{table}[h!]
\begin{center}
\centering
\begin{tabular}{|c|c|c|c|c|} 
 
 \hline

 \footnotesize Feature & \footnotesize Model & \shortstack{\footnotesize Policy}   & \footnotesize DF Eval EER\\
 \hline

  LFCC & OCS-resnet & None &  21.60  \\
 \hline
LFCC &  OCS-resnet &  SAu1 &  24.74  \\
  \hline
  LFCC &  OCS-resnet & SAv1 & \textbf{21.51}  \\
    \hline
 \hline
   LogSpec & ResNet & None &   17.03 \\
      \hline
    LogSpec & ResNet & SAu3 &    20.55 \\
   \hline
    LogSpec & ResNet & SAv3 &  \textbf{16.0} \\
    \hline
  LogSpec & ResNet & SAu4 &    16.6 \\
\hline
  LogSpec & ResNet & SAv4 &  \textbf{15.46} \\
\hline
\end{tabular}
\caption{Performance comparison: SpecAverage vs. SpecAugment on DF evaluation. According to the experiments we've conducted, SpecAverage has shown better performance.}
\label{DF_singles_SA}
\end{center}
\end{table}


\begin{figure}[h!]
\includegraphics[trim={210pt 150pt 420pt 10pt},clip,width=\columnwidth]{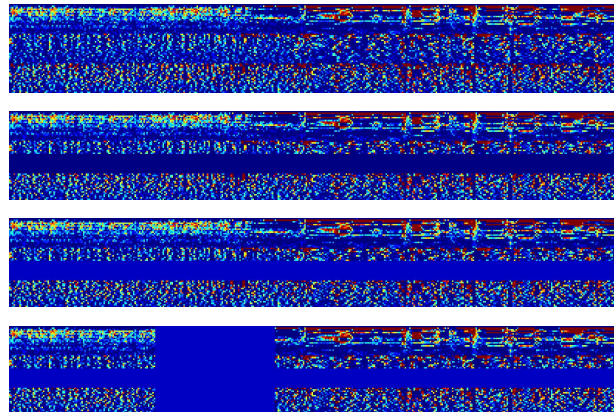} 
\caption{\it LFCC of utterances from top to bottom: No masking, SpecAugment masking in frequency, SpecAverage masking in frequency, SpecAverage masking in time and frequency.}
\label{SPECAVG}
\end{figure}

\section{Deep Fake: Results \& Analysis}
\noindent In this section we present our results for our compression robust systems. We first present the data sets used, and then the results followed by analysis. 


\subsection{Deep Fake Data Sets}
During evaluation, we aimed to create a diverse data set for training and for development. While keeping the original partition of training and development as in the ASVspoof 2019 competition, We chose the training set so that it contained low quality and high quality augmented audio with different compressions, and we chose the development set in such a way that it would contain unseen bit rates. We trained three different models with different features so we could reduce the correlation between one another and then maximize fusion results. Table \ref{table:data} contains the training and development data sets we used in the DF category.

\begin{table}[h!]
\scriptsize
\centering
\begin{tabular}{|c|c|c|c|c|c|c|c|c|c|} 
 
 \hline
\multirow{2}{*}{ Refr.} & 
\multirow{2}{*}{ Orig} & 
\multicolumn{5}{c|}{MP3} &
\multicolumn{3}{c|}{m4a}
\\
\cline{3-10}
  & & 16 & 48 & 96 & 128 & 160 & 64 & 96 & 128\\
 \hline
 Tr1&\cmark & \cmark & & & & &\cmark & &\\
 \hline
 \multicolumn{1}{|c|}{Tr2} & \multicolumn{1}{c|}{\cmark}& & \multicolumn{1}{c|}{\cmark} & & & & \multicolumn{1}{c|}{\cmark} & & \\
 \hline
 \multicolumn{1}{|c|}{Tr3} & \multicolumn{1}{c|}{\cmark} & & \multicolumn{1}{c|}{\cmark} & & & & & & \\
 \hline
 \multicolumn{1}{|c|}{Tr4} & \multicolumn{1}{c|}{\cmark} &  & & \multicolumn{1}{c|}{\cmark} & \multicolumn{1}{c|}{\cmark} & \multicolumn{1}{c|}{\cmark} & & \multicolumn{1}{c|}{\cmark} & \multicolumn{1}{c|}{\cmark} \\
  \hline
   \multicolumn{1}{|c|}{Tr5} & \multicolumn{1}{c|}{\cmark} &  & & & &  & & &   \\
  \hline
  \hline
  
     \multicolumn{1}{|c|}{Dv1} & \multicolumn{1}{c|}{\cmark} &  & \multicolumn{1}{c|}{\cmark}& \multicolumn{1}{c|}{\cmark}& \multicolumn{1}{c|}{\cmark}&  & \multicolumn{1}{c|}{\cmark}& \multicolumn{1}{c|}{\cmark}&\multicolumn{1}{c|}{\cmark}  \\
  \hline
   \multicolumn{1}{|c|}{Dv2} & \multicolumn{1}{c|}{\cmark} &\multicolumn{1}{c|}{\cmark}  &\multicolumn{1}{c|}{\cmark} & &\multicolumn{1}{c|}{\cmark} &  &\multicolumn{1}{c|}{\cmark} & &  \\
  \hline
 \end{tabular}
 \caption{Train and development data sets, DF. }
 \label{table:data}
\end{table}

\noindent Dv1 was used for the LFCC based models and Dv2 was used for the LogSpec models.


\subsection{Deep Fake Results}
\noindent Our results for single systems are presented in Table \ref{DF_singles}.


\begin{table}[h!]
\footnotesize
\centering
\centering
\begin{tabular}{|c|c|c|c|c|c|} 
 
 \hline

\footnotesize Refr. & \footnotesize Features & \footnotesize Model & \shortstack{\footnotesize Data, Policy}  & \footnotesize Norm  & \footnotesize Eval EER\\
 \hline
 \hline
 \multirow{4}{*}{\shortstack{Base-\\line}} & Audio & RawNet  & - & \xmark  & 22.38\\
 \cline{2-6}
  & LFCC & LCNN & - & \xmark  &  23.48 \\
 \cline{2-6}
  & LFCC & GMM & - & \xmark  &  25.25 \\
 \cline{2-6}
  & CQCC & GMM  & - & \xmark & 25.56 \\
 \hline
 \hline
  D1 & LFCC &  OCSresnet & Tr4 & \xmark & 28.52  \\
 \hline
 D2 & LFCC & OCSresnet & Tr4 & \cmark & 21.60  \\
 \hline
 D3 & LFCC &  OCSresnet & Tr4, SAv2 & \cmark & 21.94  \\
  \hline
 D4 & LFCC &  OCSresnet & Tr4, SAv1 & \cmark & \textbf{21.51}  \\
   \hline
 D5 & LogSpec & ResNet  & Tr1  &  \xmark  & 18.81\\
 \hline
 D6 & LogSpec & ResNet & Tr2  &  \xmark  &  18.21 \\
 \hline
 D7 & LogSpec & SENet  & Tr3 & \xmark  &  19.47 \\
 \hline
   D8 & LogSpec & ResNet & Tr1 &  \cmark  &  17.03 \\
\hline

  D9 & LogSpec & ResNet & Tr1,SAv3 &  \cmark  &  16.0 \\
\hline
  D10 & LogSpec & ResNet & Tr1,SAv4 &  \cmark  &  \textbf{15.46} \\
\hline
\hline
  T23 & - & - & - &  -  &  15.64 \\
\hline
  T20 & - & - & - &  -  &  16.05 \\
\hline
  T08 & - & - & - &  -  &  18.3 \\
\hline

\end{tabular}
\caption{Single System results, DF. The bottom three entrees are the top competition results stated in \cite{2021asvspoof}.}
\label{DF_singles}
\end{table}

\noindent It can be seen that:
\begin{enumerate}
    \item Feature normalization shows significant improvement in both models: EER reduction of 24\% in OCS-resnet, and 10\% in Resnet.
    \item SpecAverage improves the results of both models. 
    \item Our best single system, D10, used double sided Logspec, compression augmentation, feature normalization and SpecAverage. It achieves state of the art performance.
\end{enumerate}

\subsubsection{Deep Fake Fusion Scheme}
Fusion was performed using mean of the best two systems:

\begin{table}[h!]
\label{fusions_df}
\centering
\begin{tabular}{|c|c|c|} 
 
 \hline
 Single Systems & Method & Eval EER \\
 \hline
  D9+D10 & mean  & \textbf{14.46} \\
 \hline
\end{tabular}
\caption{System Fusion, DF.}
\label{DF_fusion}
\end{table}

\subsection{Deep Fake Analysis}
We used analysis provided by the competition organizers in the post evaluation phase, since the labels of the evaluation data aren't published at the time of this work. Table \ref{DF_cond} contains the DF evaluation conditions:

\begin{table}[h!]
\footnotesize
\centering
\begin{tabular}{|c|c|c|c|} 
 
 \hline
 Cond. & Compression (Quality) & VBR [kbps] & \shortstack{Double\\ Compression}\\
 \hline
DF-C1 & None & None  & \xmark\\
\hline 
DF-C2 & mp3 (low) & 80-120 & \xmark \\
\hline 
DF-C3 & mp3 (high)  & 220-260 & \xmark\\
\hline 
DF-C4 & m4a (low)  & 20-32 & \xmark\\
\hline 
DF-C5 & m4a (high) & 96-112 & \xmark\\
\hline 
DF-C6 & ogg (low)  & 80-96 & \xmark\\
\hline 
DF-C7 & ogg (high) & 256-320 & \xmark\\  
\hline 
DF-C8 & \shortstack{mp3 (low) $\rightarrow$ m4a (high)} & \shortstack{80-120, 96-112} & \cmark\\  
\hline 
DF-C9 & \shortstack{ogg (low) $\rightarrow$ m4a (high)} & \shortstack{80-96, 96-112} & \cmark\\
 
 \hline
\end{tabular}
\caption{Deep Fake evaluation data conditions. VBR denotes the variable bit rate range. DF-C8 and DF-C9 have been compressed twice.}
\label{DF_cond}
\end{table}


 

\noindent Table \ref{EER_DF} presents our system EER for each condition. Our best system outperformes the baseline systems in all conditions. DF-C4 was the condition with the worst performance. This can be explained by the fact that DF-C4 is compressed using m4a with a very low bit rate, and we assume that the low quality created this performance bias. Surprisingly, the best performance was
achieved in DF-C6, compressed using low quality ogg. This is a compression method that our model hasn't seen in training or development, emphasizing the compression robustness our system offers. Another interesting fact is that our system performed well in DF-C8.
Compressing an audio file twice using different compression methods with different bit rates enlarges the information loss, and our system performed relatively well in that sense.

\section{Logical access: Results \& Analysis}
\noindent In this section we present our results for the Logical Access category. We first present the data sets used, and then the results followed by analysis. Results are provided in terms of Equal Error Rate (EER) and min t-DCF (referred to as t-DCF), as stated in the ASVspoof 2021 evaluation plan.

\subsection{Logical Access Data Sets}
\noindent In Table \ref{LA_dataset_table} we present the data sets we used in the LA partition. For each type of feature we needed to decide what sampling rate to use, since some of the codecs include filtering and down sampling to 8 kHz (narrow band codecs) and some were wide band codecs that were in 16 kHz. For Logspec we used data that was resampled to 16 kHz no matter the codec chosen (resulting in 'half empty' spectrograms for the narrow band codecs). For LFCC, we either downsampled all of the data to 8 kHz and then performed feature extraction, or we performed effective bandwidth detection (using the obw() function in Matlab) and then trained 2 different models in order to achieve a band selective model, meaning that each new sample had its effective bandwidth calculated and then sent to the appropriate model (L7 in Table \ref{LA_singles}).

\begin{table}[h]
\centering
\renewcommand\thetable{13}
\begin{tabular}{|c|c|c|} 
 \hline
 Reference & Sample Rate[kHz] & Codecs  \\
 \hline
 1 & 8  &  all  \\
 \hline
 2 & 16 &  all  \\
 \hline
 3 & 16 &  AMR-WB, Silk-WB, G.722  \\
 \hline
\end{tabular}
\caption{Logical Access data sets.}
\label{LA_dataset_table}
\end{table}

\subsection{Logical Access Results}

Our results for single systems are presented in Table \ref{LA_singles}.
It can be seen that:
\\
\begin{table}[h]
\footnotesize
\centering
\renewcommand\thetable{14}
\begin{tabular}{|p{0.5cm}|c|c|c|c|c|c|} 
 
 \hline
Refr. & Features & Model & \shortstack{ Train \\  Set}  & \shortstack{BR\\ kHz} & \shortstack{Eval \\EER} & \shortstack{Eval \\t-DCF}\\
 \hline
 \hline
  \multirow{4}{*}{\shortstack{Base-\\line}} & Audio & RawNet  & - &  16 & 9.50 & 0.425\\
 \cline{2-7}
  & LFCC & LCNN & - &8 & 9.26 &  0.344 \\
 \cline{2-7}
  & LFCC & GMM & - & 8 & 19.30 &  0.575 \\
 \cline{2-7}
  & CQCC & GMM  & - & 8 & 15.62 & 0.497 \\
 \hline
 \hline
 L1 & LogSpec & ResNet & 2  & 16 & \textbf{5.18} & \textbf{0.293}\\
 \hline
 L2 & LogSpec & SENet  & 2 & 16 & 6.14 &  0.296 \\
 \hline
 L3 & LFCC & OCSresnet  & 1  & 8 & 7.22 &  0.333 \\
 \hline
 L4 & LFCC &  OCSresnet  & 1 & 8 & 6.65 & 0.343  \\
 \hline
 L5 & LFCC & OCSresnet  & \shortstack{1, SAu1} & 8 & 6.59 & 0.363  \\
 \hline
 L6 & LFCC &  OCSresnet  & \shortstack{1, SAv1} & 8 & \textbf{5.99} & \textbf{0.323} \\
 \hline
 L7 & LFCC &  OCSresnet & 1,3  & both & 6.99 & 0.348 \\
 \hline
\end{tabular}
\caption{Single Systems, LA.}
\label{LA_singles}
\end{table}

\begin{table*}[!ht]
\footnotesize
\centering
\renewcommand\thetable{12}
\begin{tabular}{|c|c|c|c|c|c|c|c|c|c|c|} 
 
 \hline
Refr. & DF-C1 & DF-C2 & DF-C3 & DF-C4 & DF-C5 & DF-C6 & DF-C7 & DF-C8 & DF-C9 & EER\\
 \hline
 \hline
  RawNet & 26.98 & 27.63 & 27.49 & 26.72 & 27.23 & 18.80 & 18.67 & 18.74 & 19.10 &  22.38 \\
  \hline
  LCNN & 23.19 & 34.21 & 23.88 & 25.22 &  23.85 & 19.06 & 17.10 & 28.35 & 18.54 & 23.48  \\
 \hline
  GMM LFCC & 17.39 & 39.20 & 17.97 & 20.95 & 21.43 & 22.16 & 14.83 & 39.03 & 26.65 & 25.25  \\
 \hline
  GMM CQCC & 19.48 & 48.86 & 20.37 & 19.55 & 20.27 & 17.92 & 14.42 & 49.41 & 17.39 & 25.56 \\
 \hline
 \hline

 D11+D12 &  \textbf{14.62} &	 \textbf{15.72} 	& \textbf{14.70} & \textbf{19.21} & \textbf{14.91} &	 \textbf{11.12} & \textbf{12.35} & \textbf{12.16} & \textbf{16.39} & \textbf{14.46}	  \\
 \hline
\end{tabular}
\caption{DF evaluation results on the different conditions.}
\label{EER_DF}
\end{table*}

\begin{enumerate}
    \item All of our single systems are trained with compression and transmission augmentation and perform better then the baseline systems.
    \item SpecAverage improves the results of OCS-resnet and outperformes SpecAugment. 
    \item Our best single system, L1, reduced the best baseline EER by 44\% and the best baseline min t-DCF by 15\%.
\end{enumerate}

\subsubsection{Logical Access Fusion Scheme}
In the LA task, a weighted mean based on a grid search was performed. The development set contained all channels, all codecs and all conditions.

\begin{table}[h!]

\centering
\begin{center}
\centering
\begin{tabular}{|c|c|c|c|} 
 \hline
 Single Systems & Method & Eval EER & Eval t-DCF\\
 \hline
 L1-L7 & weighted mean  & \textbf{4.66}  & \textbf{0.2882} \\
 \hline
\end{tabular}
\caption{Fusion Systems, LA.}
\label{LA_fusions}
\end{center}
\end{table}

\noindent As seen in Table \ref{LA_fusions}, the fusion scheme further improved the results. the best baseline system EER was reduced by as much as 50\%, and the min t-DCF was reduced by 16\%.

\subsection{Analysis - Logical Access}
Table \ref{LA_cond} contains the LA evaluation conditions:

\begin{table}[h]
\begin{tabular}{|c|c|c|c|} 
 
 \hline
 Cond. & Codec & Bandwidth & Transmission\\
 \hline
LA-C1 & None & 16  & None\\
\hline 
LA-C2 & a-law & 8 & VoIP \\
\hline 
LA-C3 & unk.+$\mu$-law& 8 & PSTN+VoIP\\
\hline 
LA-C4 & G.722 & 16 & VoIP\\
\hline 
LA-C5 & $\mu$-law & 8 & VoIP\\
\hline 
LA-C6 & GSM & 8 &VoIP\\
\hline 
LA-C7 & OPUS & 16 & VoIP\\  
 \hline
\end{tabular}
\caption{LA Evaluation Conditions.}
\label{LA_cond}
\end{table}

\begin{table*}[b]
\footnotesize
\centering
\begin{center}
\centering
\begin{tabular}{|c|c|c|c|c|c|c|c|c|} 
 
 \hline
Refr. &  LA-C1 & LA-C2 & LA-C3 & LA-C4 & LA-C5 & LA-C6 & LA-C7 & EER\\
 \hline
 \hline
  RawNet & 5.84 &  6.59 & 16.72 & 6.41 &  6.33 & 10.66 &  7.95 & 9.50 \\
  \hline
  LCNN &  6.71 & 8.89 & 12.02 & 6.34 &   9.25 & 11.00 &  6.66 &  9.26  \\
 \hline
  GMM LFCC & 12.72 & 21.21 & 35.55 & 15.28 & 18.76 & 18.46 & 12.73 & 19.30  \\
 \hline
  GMM CQCC & 10.57 & 14.76 & 20.58 & 11.61 & 13.58 & 14.01 & 11.21 & 15.62 \\
 \hline
 \hline
 L1-L7 &  \textbf{3.03}  & \textbf{5.04} &  \textbf{5.82} &  \textbf{3.21} &  \textbf{4.80} &  \textbf{5.82} &  \textbf{4.29}  & \textbf{4.66}\\

 \hline
\end{tabular}
\caption{LA evaluation results on the different conditions.}
\label{EER_LA}
\end{center}
\end{table*}

\noindent Aside from LA-C1, LA-C4 and LA-C6, the codecs used are ones our model has not been trained on. We can see that there are different transmission settings as well. Table \ref{EER_LA} shows the EER performance on all conditions. It can be seen that our system surpasses the baseline systems EER performance by a large margin. In addition, the performance difference between narrowband data (C2,C3,C5,C6) and wideband data (C1,C4,C7) is significant. We hypothesize that the removal of the high frequencies caused a clear degradation in performance. Finally, it can seen that our system performs relatively well even on unseen channels and codecs, as even the worst EER value (5.82) is low compare to the baseline equivalents.

\section{Discussion and insights}
In this section we will discuss interesting points and insights that we found during our research.

\subsection{The importance of augmentation}
Throughout this work, two different augmentation methods were proposed:
compression augmentation for the DF part, and codec, channel effect, and bandwidth difference augmentation for the LA part. We've shown that both methods are crucial for the models to function properly with data that has witnessed these effects or similar ones. Both augmentation types not only help models deal with the trained data, but allow the models to be robust to different kinds of compressions and channels. The evidence of this is clear, as our best DF system has the best result on an unseen compression (ogg), and our best LA system has good results on unseen channels.    


\subsection{SpecAverage vs. SpecAugment}
Throughout this work, two different online masking methods were considered. Based on the tests that we've performed we hypothesize that the reason SpecAverage had better performance is that the average value is statistically meaningful relative to the input feature, offering more information in addition to the regularization effect provided by the masking. It is important to note that time warping, one of the features in SpecAugment, hasn't been used as it is costly in computing resources. Despite that fact, masking with the average value gave consistently better results than masking with the value 0.

\subsection{LFCC vs. Log Spectrogram}
During this work we saw that the LFCC based model underperformed both of the Log Spectrogram models. We believe that a possible explanation for this might be in the dynamic LFCC features. As we visualized in Figure \ref{LFCC_compression}, the differences between the same audio file uncompressed or compressed with different bitrates can be clearly seen, especially in the high frequency range. While the Log Spectrogram contains all of the information as it comes, we believe that an additional amount of noise is produced during the LFCC extraction process while computing the high frequency derivatives. The differences in the dynamic features between bitrates can be explained by the fact that lossy compression of an audio signal involves removing high frequencies that are unheard by the human ear, and the lower the quality the more information is lost, together with harsher quantization. A future work solution can be to try and use only parts of the dynamic LFCC features, or even remove them.  

\subsection{Dealing with unseen and double compressions}
In the DF part, we've encountered an interesting phenomenon: our model that has been trained on MP3 and m4a compressions performed best on low quality ogg, which is a compression method that hasn't been learned by the model. We believe a reason for that can be that although the compression methods are different, they do remove specific frequencies that are based on human hearing perception tests, and in that manner they are similar. We believe that this fact contributed to this result. To further test this interesting fact we conducted an experiment, as seen in Table \ref{compression_robustness_test}. The results were consistent.

\begin{table}[h!]
\centering
\begin{center}
\centering
\begin{tabular}{|c|c|c|c|} 
 
 \hline
 System & \shortstack{Compressions \\ trained on} & \shortstack{Compression \\ tested on } & EER \\
 \hline
  D4 & MP3, M4a  & MP3, M4a & 3.63 \\
 \hline
 D4 & MP3, M4a  & Opus & 3.7 \\
 \hline
 D4 & MP3, M4a  & ogg & 3.82 \\
 \hline
\end{tabular}
\caption{Compression robustness test. Our model performes well even on unseen compressions. The test data set used for comparison was the ASVspoof 2019 LA evaluation set.}
\label{compression_robustness_test}
\end{center}
\end{table}

\subsection{The bitrate effect}
During the evaluation, we experimented with a large variety of bitrates. One of the challenges was to decide which bitrates to use for training and development for each compression method and which not to use. In order to further understand how to decide, we performed internal tests and came up with a few conclusions:

\begin{enumerate}
    \item Knowing the exact bitrates in the test set gives the best performance (given that they are in fact known).
    \item If the bitrates in the test set aren't known, the best performance was achieved by using at least one low bitrate and at least one high bitrate. The trade off of using more is time (training and testing), and doesn't always help.
    \item The development set used in the training process should contain bitrates unseen during training.
\end{enumerate}

\subsection{Out of domain data, DF part}
Although our system has good relative performance both to the baseline systems and to the other teams that competed in the competition, the overall EER is not satisfying for a real life system. We believe that the reason for this is the limitation to only use ASVspoof 2019 data. As presented in the ASVspoof 2021 workshop, the DF evaluation dataset contains data from the VCC 2018 \cite{VCC2018} and VCC 2020 \cite{VCC2020} contests. In order to maintain good results and updated systems, new datasets containing a large variety of distributions must be used in the training process of the models.




\section{Conclusions and Future Work}
In this paper we performed an in depth study of how data augmentation affects voice anti-spoofing systems. We presented
two different types of data augmentation (for DF and for LA),
that both significantly increased the results of the models we
used. Our methods showed improvement in tasks that involve new spoofing attacks that haven’t been seen during training, compressed data, transmitted
data and data with different bandwidths. We introduced two
new methods of online augmentation - SpecAverage and feature
normalization, that both contributed to our results. Our feature
design that included double sided Logspec helped increase the
results as well. The combination of our methods achieved state of the art results in the DF category, both with a single system and with a system fusion. Given that our methods are mostly used on the audio frequency based features, we believe that they can be used in other audio related tasks such as speaker recognition among others. 

\section{Declaration of competing interest}
The authors declare that they have no known competing financial interests or personal relationships that could have appeared to influence the work presented in this paper.

\bibliographystyle{IEEEbib}
\bibliography{ASVspoof2021_bibliography}

\end{document}